# High-Order Perfect Absorption in the Absence of Exceptional Point

Huisheng Xu,[1] Luojia Wang,[2] Luqi Yuan[2,*] and Liang Jin[1,†]

[1]*School of Physics, Nankai University, Tianjin 300071, China*
[2]*State Key Laboratory of Photonics and Communications, School of Physics and Astronomy, Shanghai Jiao Tong University, Shanghai, 200240, China*

High-order perfect absorption of coherent input has recently attracted significant attention due to its broadband absorption capacity. However, the realization of a high-order perfect absorber relies on the exceptional point (EP) to coalesce the scattering zeros. Here, we present a general scattering framework and achieve the high-order perfect absorber in the absence of an EP. We consider the asynchronous coherent input, where a spatial delay introduces a momentum-dependent phase factor beyond the amplitude and phase control in synchronous coherent input. This new degree of freedom enables active control of the momentum-dependent output, effectively reshaping the absorption line shape necessary for the high-order perfect absorber. Remarkably, despite the absence of an EP, the proposed high-order perfect absorber exhibits significant response to the perturbations in the delay length. Our findings provide insights for the delay-induced momentum-sensitive interference phenomenon and offer a new route for wave control.

*Introduction*—Coherent perfect absorption (CPA) is a hallmark non-Hermitian interference phenomenon [1–8]. When properly superposed input waves from different ports arrive simultaneously at a scattering center, they can be completely absorbed by a lossy medium. CPA has been extensively explored across diverse platforms, stimulating applications ranging from photocurrent enhancement to optical data processing [9,10]. However, CPA typically has a narrow bandwidth, where a slight frequency shift can destroy the resonance and limit its utility. To overcome this limitation, exceptional points (EPs) have been incorporated into CPA [11–13]. When two or more purely incoming wave solutions coalesce at a real frequency, the resulting high-order perfect absorber, known as a CPA EP, exhibits a broadened absorption line shape. This phenomenon has been predicted and observed in optical [14–16], electronic [17], and acoustic systems [18], offering strong potential for broadband absorption. Moreover, this high-order perfect absorber inherits the sensitivity of EPs, enabling EP-based sensing applications [19–21].

Fundamentally, the broadband absorption of high-order perfect absorbers corresponds to a high-order zero of the scattering output and does not necessitate the presence of EPs. Notably, the output after scattering depends on both the scattering center and the incident waves. In conventional CPA [1–3], the incident waves are always *synchronous* and controlled solely by their amplitudes and phases. Such input corresponds to a zero-eigenvalue eigenmode of the scattering matrix [9], thereby revealing only the properties of the scattering center. This raises the fundamental question of whether there are additional degrees of freedom in the incident waves that can be exploited to independently manipulate the scattering output. Addressing this question is essential for a comprehensive understanding of the interplay between the scattering center and the incident waves, and for realizing high-order perfect absorbers within a more general framework.

In this Letter, we establish a generalized scattering framework for *asynchronous* coherent input and demonstrate the high-order perfect absorption in the absence of EPs. The asynchronous coherent input introduces a spatial delay as a new degree of freedom for manipulating wave interference. The incorporated delay brings a momentum-dependent dynamical phase factor, enabling active control of the scattering output through the incident waves, which cannot be achieved by conventional amplitude and phase modulation. Using this mechanism, we design a high-order perfect absorber with sextic absorption line shape, which arises from the delay-induced destructive interference instead of the coalescence of scattering zeros. Since the resulting high-order absorber relies on the delay rather than an EP, it exhibits a strong absorption response to variations in delay length without spectral splitting from perturbations in system parameters. Our delay framework highlights the asynchronous wave interference in the non-Hermitian

[*]Contact author: yuanluqi@sjtu.edu.cn
[†]Contact author: jinliang@nankai.edu.cn

scattering and opens new opportunities for wave manipulation and coherent control.

*Delay formalism*—We briefly review the concept of Wigner delay [22,23], a fundamental scattering phenomenon underlying the characterization of asynchronous coherent input. Intuitively, a wave packet would reflect directly at the boundary of a potential barrier following a geometric path. However, the actual reflection involves a phase gradient across the momentum of the wave packet, shifting its center by a delay length $L$, as if it experienced a spatial delay relative to the ideal path. This delay length is quantified as

$$L = d\arg[r(k)]/dk, \quad (1)$$

where $r(k)$ is the reflection coefficient and $k$ is the momentum of the wave packet. The Wigner delay has been extensively studied in optical fiber [24], acoustic system [25], time-varying media [26], and synthetic lattice [27], with applications in precise measurement [28] and interferometry [29].

We address the inverse problem of what the influence of delay on the interference phenomenon is if it is introduced to the coherent input. We simply consider a two-port scattering system. In contrast to the Wigner delay that characterizes delay after scattering, we adapt this framework to describe relative delay in coherent input waves before scattering. The input waves in the two ports are represented as $a = (a_1, a_2)^T$ with $a_1 : a_2 = 1 : \eta$. Conventionally, the two components arrive simultaneously at the scattering center without any relative delay, and $\eta = |\eta|e^{i\varphi}$ describes their amplitude ratio and phase difference [Fig. 1(a)]. Here, we consider the input wave in port 2 having a delayed arrival with length $l$ along the propagation direction [Fig. 1(b)]. Inspired by the characterization of delay in Eq. (1), the delay length $l$ relates to the phase of $\eta$ through its gradient $l = d\arg(\eta)/dk$. Given that $l$ is a $k$ independent constant, we obtain $\eta$ after an integration,

$$\eta = |\eta|e^{i\varphi}e^{ikl}. \quad (2)$$

Notably, the delay introduces an additional momentum-dependent dynamical phase factor $e^{ikl}$.

We further consider the impact of delay on the output. Applying the scattering matrix to the input wave vector, the output wave vector is obtained $b = (b_1, b_2)^T = S(k)a$, where $S(k)$ is the scattering matrix. The momentum-dependent output for a renormalized input $a = 1/\sqrt{1+|\eta|^2}(1, |\eta|e^{i\varphi}e^{ikl})^T$ in Eq. (2) is

$$\begin{pmatrix} b_1 \\ b_2 \end{pmatrix} = \frac{1}{\sqrt{1+|\eta|^2}} \begin{pmatrix} s_{11}(k) + s_{12}(k)|\eta|e^{i\varphi}e^{ikl} \\ s_{21}(k) + s_{22}(k)|\eta|e^{i\varphi}e^{ikl} \end{pmatrix}, \quad (3)$$

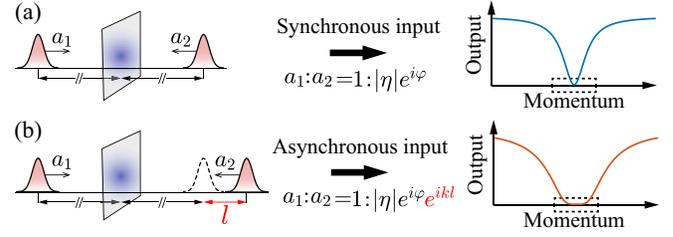

FIG. 1. (a) Coherent input without relative delay, where the coherent control only involves amplitude ratio and phase difference. (b) Coherent input with relative delay, where the delay length $l$ causes an extra momentum-dependent phase $e^{ikl}$. Right panels: delay-induced modification of the momentum-dependent output.

where $s_{qp}(k)$ is the scattering coefficient describing the output in port $q$ for the input in port $p$.

Notably, the dependence of output on the momentum $k$ stems from both the scattering coefficients $s_{qp}(k)$ and the delay-induced dynamical phase $e^{ikl}$. Therefore, the delay provides an additional control parameter that actively manipulates the momentum dependence of the output. This leads to novel scattering phenomena that are absent under synchronous coherent input. For example, the delay facilitates a momentum-sensitive modulation of absorption spectrum, where tuning the delay length $l$ can enhance the absorption without relying on an EP. To illustrate this effect, we consider the derivatives of the output components $b_1$ and $b_2$ with respect to the momentum $k$. In the context of perfect absorption $b_1 = b_2 = 0$, the derivatives are

$$\frac{d}{dk}b_1 = \frac{s_{11}(k)}{\sqrt{1+|\eta|^2}}\left[\frac{d}{dk}\ln\frac{s_{11}(k)}{s_{12}(k)} - il\right], \quad (4)$$

$$\frac{d}{dk}b_2 = \frac{s_{21}(k)}{\sqrt{1+|\eta|^2}}\left[\frac{d}{dk}\ln\frac{s_{21}(k)}{s_{22}(k)} - il\right]. \quad (5)$$

By contrast to the first-order perfect absorber with only $b_{1(2)} = 0$ [Fig. 1(a)], the derivative $db_{1(2)}/dk = 0$ corresponds to the second-order perfect absorber with a broadened absorption line shape [Fig. 1(b)]. Therefore, high-order perfect absorbers are enabled by the delay through tuning the derivatives to zero [30], without the need for an EP.

In the following, we propose high-order perfect absorbers based on the delay framework and demonstrate the crucial role of delay in the wave interference.

*Synthetic frequency lattice*—CPA offers a powerful tool for rectifying energy flow. However, in discrete lattices, the absorption of an initial excitation is incomplete, as the excitation inevitably spreads in momentum space while the absorption bandwidth remains narrow [17,31]. This limitation becomes particularly severe for excitations with compact spatial profiles constrained by the lattice size [32–35]. Recently, synthetic frequency lattices offer a promising way to overcome this lattice-size constraint [36–38], with frequency modes effectively serving as discrete lattice sites [39–41].

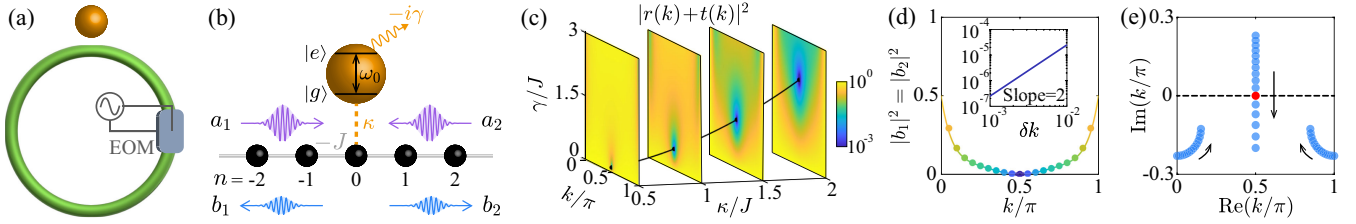

FIG. 2. (a) Schematic illustration of a ring resonator coupled to a two-level dissipative atom. The ring resonator is dynamically modulated via an electro-optic modulator (EOM). (b) The synthetic frequency lattice for (a). (c) Eigenvalue intensity $|r(k) + t(k)|^2$ of the scattering matrix. The black line marks the zeros of the eigenvalue. (d) The normalized output $|b_1|^2$ and $|b_2|^2$ versus $k$ for $\gamma = \kappa = 2J$. The inset displays the double-logarithmic plot near $k = \pi/2$, with $\delta k = k - \pi/2$. (e) Trajectories of three purely incoming wave solutions at fixed $\kappa = 2J$ as $\gamma/J$ varying from 0 to 3.

Figure 2(a) shows a design of the dynamically modulated ring resonator coupled to a dissipative two-level atom [42–44]. An electro-optic modulator, driven by a sinusoidal external source $-2J\cos(\Omega t)$, is placed within the resonator. The dynamic refractive index modulation induces coupling between adjacent resonant modes $\omega_n = \omega_0 + n\Omega$, and creates a uniform synthetic frequency lattice [45–47]. The two-level atom with ground state $|g\rangle$ and excited state $|e\rangle$ is separated by the frequency $\omega_0$, so the transition between $|g\rangle$ and $|e\rangle$ is resonantly coupled to the zeroth resonant mode with the coupling strength $\kappa$ [48]. The atom also possesses an intrinsic dissipation rate $\gamma$. Such a two-level atom can be implemented using a Cs atom coupled to a whispering-gallery-mode resonator [49], or alternatively in superconducting platforms using artificial atoms, such as superconducting qubits coupled to transmission-line resonators [43,44]. The Hamiltonian of this model is

$$H = \omega_n \sum_n \hat{c}_n^\dagger \hat{c}_n - 2J\cos(\Omega t) \sum_n (\hat{c}_n^\dagger \hat{c}_{n+1} + \hat{c}_{n+1}^\dagger \hat{c}_n)$$
$$+ (\omega_0 - i\gamma)|e\rangle\langle e| + \kappa(\hat{c}_0^\dagger|g\rangle\langle e| + \hat{c}_0|e\rangle\langle g|), \quad (6)$$

where $\hat{c}_n^\dagger$ ($\hat{c}_n$) is the creation (annihilation) operator for the $n$th resonant mode.

Under rotating wave approximation, the above Hamiltonian describes a uniform tight-binding chain side-coupled with a dissipation [Fig. 2(b)]. The dynamics of photon transport is characterized by the equations of motion for the single excitation (Supplemental Material Sec. A [50]),

$$i\frac{df_n}{dt} = \omega_0 f_n - J f_{n-1} - J f_{n+1}, \quad (|n| > 0), \quad (7)$$

$$i\frac{df_0}{dt} = \omega_0 f_0 - J f_{-1} - J f_1 + \kappa f_\gamma, \quad (8)$$

$$i\frac{df_\gamma}{dt} = (\omega_0 - i\gamma) f_\gamma + \kappa f_0, \quad (9)$$

where $f_n$ describes the wave amplitude of the photon in the $n$th lattice site while the atom is in the ground state, $|g\rangle$, and $f_\gamma$ denotes the wave amplitude of the atom being excited to $|e\rangle$ by the propagating photon. We emphasize that the above equations of motion are generic and can be realized in various realistic platforms beyond synthetic dimensions (Supplemental Material Sec. B [50]).

In the elastic scattering process, the wave amplitudes have the form of $f_n = \psi_n e^{-i\omega t}$, where $\psi_n$ is the steady-state wave amplitude. As the system has the inversion symmetry, the reflection and transmission are symmetric [60], with $s_{11}(k) = s_{22}(k) = r(k)$ and $s_{21}(k) = s_{12}(k) = t(k)$, being independent of the input direction. By setting $\psi_n = e^{ikn} + r(k)e^{-ikn}$ for $n \leq 0$ and $\psi_n = t(k)e^{ikn}$ for $n \geq 0$, the scattering coefficients are obtained (Supplemental Material Sec. C [50]),

$$r(k) = t(k) - 1 = -\frac{\kappa^2}{2J\gamma\sin(k) + 2iJ^2\sin(2k) + \kappa^2}. \quad (10)$$

The corresponding scattering matrix $S(k)$ is nonunitary due to the dissipation [61–63]. The eigenvalues of $S(k)$ consist of a momentum-dependent component $r(k) + t(k)$ and a constant value of $-1$. Figure 2(c) plots $|r(k) + t(k)|^2$ as a function of momentum $k$. At the resonant input $k = \pi/2$ for $\kappa^2 = 2J\gamma$, the eigenvalue reaches zero, as indicated by the black line. Notably, a zero eigenvalue of the scattering matrix identifies a perfect absorption [9]. The associated eigenvector is $a = 1/\sqrt{2}(1, 1)^T$, leading to $b = S(k)a = 0$. This indicates that the resonant waves synchronously input from opposite directions with identical amplitude and phase are perfectly absorbed by the atom (Supplemental Material Sec. D [50]).

Without loss of generality, we show the normalized output $|b_1|^2$ and $|b_2|^2$ as a function of momentum $k$ for $\gamma = \kappa = 2J$ in Fig. 2(d). The perfect absorption displays a quadratic line shape, as evidenced by a slope of 2 in the logarithmic plot [the inset of Fig. 2(d)]. The quadratic line shape suggests that the perfect absorption is first-order and occurs without the coalescence of scattering zeros [11]. We analytically continue $k$ to the complex plane and solve for the purely incoming wave solutions [64,65]. Figure 2(e) displays three effective complex solutions for $k$ (Supplemental Material Sec. E [50]), only one of which moves to the real axis (red dot) at $\gamma = \kappa = 2J$.

*High-order perfect absorption*—The perfect absorption shown in Fig. 2(d) exhibits a narrow bandwidth, that is, the output wave near $k = \pi/2$ deviates from zero rapidly. A broadband absorption spectrum of high-order perfect absorber resolves this issue [14]. We utilize our delay formalism to create the high-order perfect absorption with an EP being absent.

From the delay formalism, the outputs are obtained by substituting Eq. (10) into Eq. (3). Perfect absorption occurs when the coherent input waves have equal amplitude $|\eta| = 1$ and satisfy the phase-matching condition $e^{i\varphi}e^{ikl} = 1$. Compared with the scheme without delay [14–21], the phase matching is now jointly tuned by the relative phase $\varphi$ and the delay length $l$. Notably, the output derivatives can be tuned to zero via the delay, allowing the active reshaping of the absorption line shape to realize high-order perfect absorption. Under perfect absorption $b_1 = b_2 = 0$, the output derivatives are shown in Figs. 3(a) and 3(b). Remarkably, when $l = 2J/\gamma$ and $l = -2J/\gamma$, the derivatives vanish ($db_1/dk = 0$ and $db_2/dk = 0$), giving rise to second-order perfect absorption in ports 1 and 2, respectively. Furthermore, in the case $\gamma = \kappa = 2J$, an even higher-order perfect absorption is possible. Specifically, when $l = 1$ and $l = -1$, the second derivatives also vanish ($d^2b_1/dk^2 = 0$ and $d^2b_2/dk^2 = 0$), leading to a third-order perfect absorption in ports 1 and 2. As shown in Fig. 3(c), the absorption bandwidth varies with delay and system parameters. Increasing the order of perfect absorption via tuning the delay results in an ultrawide absorption bandwidth around $k = \pi/2$, where the output approaches zero over a broad momentum range. Figures 3(d) and 3(e) show the outputs for ports 1 and 2 with delay length $l = 1$. Unlike the delay-free case in Fig. 2(d), where $|b_1|^2$ and $|b_2|^2$ are identical, the introduction of delay induces third-order perfect absorption in port 1 and first-order perfect absorption in port 2. Notably, the broadened absorption bandwidth can also be characterized in terms of the incident frequency, with the absorption order identical to that extracted from the incident momentum [66].

The proposed high-order perfect absorption arises from the actively introduced delay in the coherent input, which fundamentally differs from that achieved from the coalescence of scattering zeros (Supplemental Material Sec. F [50]). The delay scheme highlights the interplay between the momentum dependences of scattering coefficients and the input. By contrast, conventional scheme from the coalescence of scattering zeros solely depends on the scattering center [11–13].

To verify the proposed high-order perfect absorption, we perform the time evolution of coherent input with a relative delay. The initial excitation is a superposition of two counterpropagating Gaussian wave packets,

$$|\phi\rangle = \frac{1}{\sqrt{1+|\eta|^2}}(|\phi_-\rangle + |\eta|e^{i\varphi}|\phi_+\rangle), \quad (11)$$

where $|\phi_\pm\rangle = \sum_n e^{-(n-n_\pm)^2/(2\sigma^2)} e^{\mp ik_c(n-n_\pm)}/\sqrt[4]{\pi\sigma^2}|n\rangle$ [67–70]. The wave packets $|\phi_-\rangle$ and $|\phi_+\rangle$ are centered at the modes $n_-$ and $n_+$ with relative amplitude $|\eta|$, phase $e^{i\varphi}$, and delay length $l = |n_+| - |n_-|$. Here, $\sigma$ controls their width and $k_c$ is the central momentum. This excitation can be implemented in the synthetic frequency lattice by temporally shaping the intraresonator field within a single round trip, where the temporal position of the injected pulse determines the momentum (Supplemental Material Sec. G [50]). In Fig. 3(f), the two wave packets are shown to be nearly perfectly absorbed by the dissipative atom. The incomplete absorption is attributed to the spreading of Gaussian wave packets near $k_c = \pi/2$ in the momentum space.

The distinct outputs in ports 1 and 2 shown in Fig. 3(f) imply different orders of perfect absorption. To quantify this difference, we calculate the residual intensity of the wave packets after scattering in ports 1 and 2, denoted as $I_1$ and $I_2$, respectively. Through analytical derivation (Supplemental Material Sec. H [50]), we find that the residual intensity of the wave packets after the $n$th-order perfect absorption scales with the wave packet width as

$$I_s \propto \sigma^{-2n}, \quad (12)$$

where the subscript $s = 1, 2$ denotes the port. In Fig. 4(a), we plot the residual intensities at both ports for varying packet widths $\sigma$ [71]. The scaling laws confirm the third-order perfect absorption in port 1 and the first-order perfect absorption in port 2. We highlight that the enhanced performance of high-order perfect absorption originates from its ultrawide absorption bandwidth.

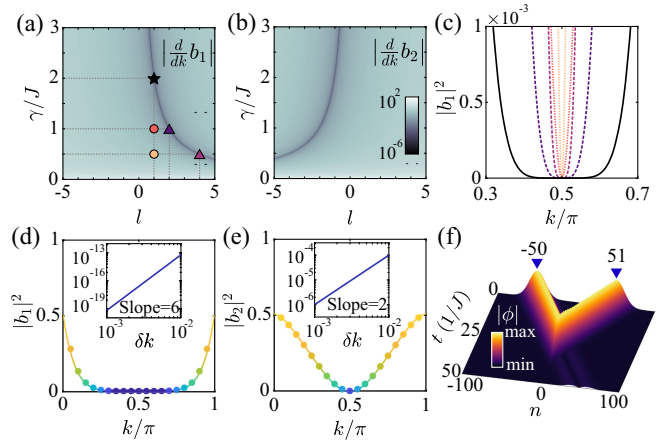

FIG. 3. (a),(b) Derivatives of $b_1$ and $b_2$ under perfect absorption $b_1 = b_2 = 0$. (c) Comparison of absorption line shapes $|b_1|^2$ near $k = \pi/2$ for a set of parameters marked in (a). (d),(e) Output intensities for ports 1 and 2, exhibiting third-order (slope 6) and first-order (slope 2) absorption for $\gamma = \kappa = 2J$ and $\lambda = 1$. (f) Time evolution of the amplitude $|\phi|$ for two coherent input wave packets, initially centered at $n_- = -50$ and $n_+ = 51$, with $\sigma = 10$.

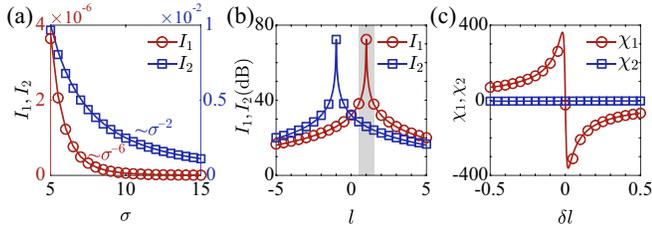

FIG. 4. (a) Scaling behaviors of residual intensity for third-order and first-order absorption. (b) Absorption in decibels for ports 1 and 2 as a function of delay $l$, exhibiting peaks at $l = 1$ and $l = -1$, respectively. (c) $\chi_1$ and $\chi_2$ versus the delay perturbation $\delta l$ near $l = 1$ [gray region in (a)].

*Delay modulation*—In the EP-based schemes [72–77], achieving high-order perfect absorption requires tuning the system parameters to reach the EPs, and tiny parameter variations near the EP induce large splittings in the absorption spectrum [19–21]. By contrast, the delay-based scheme does not work at an EP, and the perturbations of system parameters do not lead to spectrum splitting [Fig. 2(c)]. Instead, our proposed high-order perfect absorption depends on the delay, and small variations in the delay can induce a strong absorption response without relying on the EP mechanism [78,79].

We consider the variation of absorption under different delays. To clearly present the variations, we express the residual intensity in decibels as $-10\log_{10}[I_{1(2)}]$. In Fig. 4(b), we plot the absorptions in ports 1 and 2 as functions of the delay length $l$ under the condition of perfect absorption. Peaks in $I_1$ at $l = 1$ and $I_2$ at $l = -1$ are observed. This rapid intensity decrease indicates the delay-induced significant enhancement of absorption. We define the $\chi_s$ as the change in absorption (in decibels) with respect to the perturbation in the delay,

$$\chi_s = \frac{-10\log_{10}(I'_s/I_s)}{\delta l}, \quad (13)$$

where $I_s$ ($I'_s$) denotes the residual intensity before (after) introducing the perturbation $\delta l$. Figure 4(c) shows $\chi_s$ as a function of $\delta l$ near the absorption peak at $l = 1$. A significant increase in $\chi_s$ near the third-order perfect absorption is clearly observed, and the enhancement originates from the sharp transition near the high-order perfect absorption peak in the scaling residual intensity from $\sigma^{-2}$ to $\sigma^{-6}$. The change in the scaling exponent introduces a dominant term $20\Delta n \log_{10}\sigma/\delta l$ in $\chi_s$, where $\Delta n$ is the change in the order of perfect absorption (Supplemental Material Sec. I [50]). Therefore, the modulation of delay alters the order of perfect absorption, and induces a significant change in the absorption. Moreover, the response can be further enlarged by increasing the wave packet width. The strong absorption response has the potential for sensing applications, but a full assessment would take noise into account (Supplemental Material Sec. J [50]). Further systematic analysis on the signal-to-noise ratio has profound meaning [80–82].

*Conclusion*—Traditionally, coherent input assumes synchronous excitation, with control confined to the amplitude and phase. Here, we introduce the concept of asynchronous coherent input and establish a generalized scattering framework. We show that delay, as a new degree of freedom, reshapes the absorption line shape, enables ultra-broadband perfect absorption, and enhances the response to perturbations. The broader role of delay in coherent input overturns the limitation that dissipative systems without CPA EPs cannot achieve broadband absorption, and provides a new paradigm for designing high-order absorbers. Our findings offer valuable insights for the asynchronous wave interference engineering, paving the way for designing advanced photonic devices with tailored absorption profiles. The delay mechanism also enables flexible control of reflection and transmission line shapes [83–86], and can be generalized to multiport systems [20,87,88]. While this Letter focuses on spatial delay, the underlying principles are directly applicable to temporal delay [27,89,90]. Furthermore, the delay may stimulate intriguing interference effects and transport phenomena for giant atoms coupled to multimode waveguides [43].

*Acknowledgments*—This work is supported by the National Natural Science Foundation of China (Grant No. 12525502, No. 124B2079, No. 12475021, and No. 12204304). L. J. also acknowledge the support from Quantum Science and Technology-National Science and Technology Major Project (Grant No. 2024ZD0301000). L. Y. also acknowledge the support from National Key R&D Program of China (Grant No. 2023YFA1407200).

*Data availability*—The data that support the findings of this article are not publicly available. The data are available from the authors upon reasonable request.